# Magnetic light

*Arseniy I. Kuznetsov [1#], Andrey E. Miroshnichenko[2], Yuan Hsing Fu[1], JingBo Zhang[1],*

*and Boris Luk'yanchuk[1]*

1. Data Storage Institute, 5 Engineering Drive I, 117608, Singapore

2. Nonlinear Physics Centre, Centre for Ultrahigh-bandwidth Devices for Optical Systems (CUDOS), Research School of Physics and Engineering,

Australian National University, Canberra, 0200, Australia

**Abstract**

Spherical silicon nanoparticles with sizes of a few hundreds of nanometers represent a unique optical system. According to theoretical predictions based on Mie theory they can exhibit strong magnetic resonances in the visible spectral range. The basic mechanism of excitation of such modes inside the nanoparticles is very similar to that of split-ring resonators, but with one important difference that silicon nanoparticles have much smaller losses and are able to shift the magnetic resonance wavelength down to visible frequencies. We experimentally demonstrate for the first time that these nanoparticles have strong magnetic dipole resonance, which can be continuously tuned throughout the whole visible spectrum varying particle size and visually observed by means of dark-field optical microscopy. These optical systems open up new perspectives for fabrication of low-loss optical metamaterials and nanophotonic devices.

---

[#] corresponding author. E-mail: Arseniy_K@dsi.a-star.edu.sg

It is well known that a pair of oscillating electric charges of opposite signs, oscillating electric dipole, produces electromagnetic radiation at a frequency of the oscillations [1]. Although, distinct "magnetic charges", or monopoles, have not been observed so far, magnetic dipoles are very common sources of magnetic field in nature. The field of the magnetic dipole is usually calculated as the limit of a current loop shrinking to a point. The field profile in this case is similar to that of an electric dipole with one important difference that electric and magnetic field are exchanged. The most common example of a magnetic dipole radiation is an electromagnetic wave produced by an excited metal split-ring resonator (SRR), which is a basic constituting element of metamaterials (Fig.1a) [2-16]. The real currents excited by external electromagnetic radiation and running inside the SRR produce a transverse oscillating up and down magnetic field in the center of the ring, which simulates an oscillating magnetic dipole. The major interest to these artificial systems is due to their ability to response to a magnetic component of incoming radiation and thus to have a non-unity or even negative magnetic permeability ($\mu$) at optical frequencies, which does not exist in nature. This provides possibilities to design unusual material properties such as negative refraction [2-16], cloaking [17, 18], or superlensing [19]. The SRR concept works very well for gigahertz [8-10], terahertz [11, 12] and even near-infrared (few hundreds THz) [13-16] frequencies. However, for shorter wavelengths and in particular for visible spectral range this concept fails due to increasing losses and technological difficulties to fabricate smaller and smaller constituting split-ring elements [14, 20]. Several other designs based on metal nanostructures have been proposed to shift the magnetic resonance wavelength to the visible spectral range [2-6, 21-25]. However, all of them are suffering from losses inherent to metals at visible frequencies.

An alternative approach to achieve strong magnetic response with low losses is to use nanoparticles made of high-refractive index dielectric materials [6, 26]. As it follows from the exact Mie solution of light scattering by a spherical particle, there is a particular parameter range where strong magnetic dipole resonance can be achieved. Remarkably, for the refractive indices above a certain value there is a well-established hierarchy of magnetic and electric resonances (see Supplementary Figure 1). In contrast to plasmonic particles the first resonance of dielectric nanoparticles is a magnetic dipole resonance, and takes place when the wavelength of light inside the particle equals to the diameter $\lambda_0 / n_{sp} \approx 2R_{sp}$. Under this condition the polarization of the electric field is anti-parallel at opposite boundaries of the sphere, which gives rise to strong coupling to circulation displacement currents while magnetic field oscillates up and down in the middle (Fig.1b).

The first experimental demonstration of magnetic response of dielectric particles and negative-index dielectric-based metamaterials has been done at gigahertz frequencies, where materials with extremely high refractive index up to several tens of refractive index units (RIU) exist [27, 28]. Several theoretical works have also discussed various realistic materials to achieve magnetic response of dielectric particles at terahertz [29-33], near-infrared [29, 34, 35], and what is more important for us, at visible frequencies [33, 36-39]. Recent experimental demonstrations also include magnetic response of single silicon carbide microrods [40] and arrays of tellurium microcubes [41], both in the midinfrared spectral range. However, the experimental proof of this concept at visible frequencies is still lacking.

"Seeing is believing" is an idiom supported by people since ancient times. To see a medium response to the magnetic field component of light, "magnetic light", by naked eyes would be an additional proof of the concept of metamaterials, which people can design to control light at the new level beyond nature.

In this paper, we experimentally demonstrate for the first time that spherical silicon nanoparticles with sizes in the range from 100 nm to 200 nm have strong magnetic dipole response in the visible spectral range. The scattered "magnetic" light by these nanoparticles is so strong that it can be easily seen under a dark-field optical microscope. The wavelength of this magnetic resonance can be tuned throughout the whole visible spectral range from violet to red by just changing the nanoparticle size.

**Results and discussion**

We choose silicon (Si) as a material which has high refractive index in the visible spectral range (above 3.8 at 633 nm) on the one side and still almost no dissipation losses on the other. Silicon nanorods have attracted considerable attention during the last few years due to their ability to change their visible color with the size [42, 43]. This effect appears due to excitation of particular modes inside the cylindrical silicon nanoresonators. Moreover, recent theoretical work predicted that spherical silicon nanoparticles with sizes in the range from 100 nm to 200 nm should have both strong magnetic and electric dipole resonances in the visible spectral range [37]. But how to fabricate silicon spheres of this size? Lithography cannot be directly applied to produce spherical particles [41, 44]. Existing bottom-up methods such as chemical or plasma synthesis typically produce very small silicon particles or a broad particle size distribution [45, 46]. The particles of different sizes are mixed together which complicate spectral analysis. In this paper, we employ laser ablation technique, which is an efficient method to produce nanoparticles of various materials and sizes [47-49]. Nanoparticles produced by the ablation method can be localized on a substrate and measured separately from each other using single nanoparticle spectroscopy.

Dark-field microscopic image of a silicon sample ablated by a focused femtosecond laser beam is shown in Fig.2a. It shines by all the colours of the rainbow from violet to red. To clarify the origin of this strong scattering we selected some nanoobjects shining with different colours on the sample (see Fig.2a) and measured their scattering spectra by single nanoparticle dark-field spectroscopy. Then, the same sample area was characterized by scanning electron microscopy and the selected nanoobjects providing different colours have been identified (Fig.2b, the dark-field microscope image is inverted in horizontal direction relative to that of the SEM). The results of this comparative analysis of the same nanoobjects by dark-field optical microscopy, dark-field scattering spectroscopy, and scanning electron microscopy are presented in Fig.3. As it can be seen from the SEM images the observed colours are provided by silicon nanoparticles of almost perfect spherical shape and varied sizes. This makes it possible to analyze scattering properties of these nanoparticles in the frames of Mie theory [50] and identify the nature of optical resonances observed in our spectral measurements. The bottom panels (iv) in Fig.3 represent a total extinction cross-section calculated using Mie theory [50] for silicon nanoparticles of different sizes (the calculations were done in free space). In these calculations, the size of the nanoparticles in each figure was chosen to be similar to the size defined from each corresponding SEM image (ii). It can be seen that there is a clear correlation between the experimental (iii) and theoretical spectra (iv) both in the number and position of the observed resonances. This makes it obvious that Mie theory describes more or less accurately our experimental results.

One of the main advantages of the analytical Mie solution compared to other computational methods is its ability to split the observed spectra into separate contributions of different multipole modes and have a clear picture of the field distribution inside the particle at each resonance maximum. This analysis was done for each particle size in Fig.3 and corresponding multipole contributions were identified (see notations in the experimental and theoretical spectra). According to this analysis the first strongest resonance of these nanoparticles appearing in the longer wavelength part of the spectrum corresponds to magnetic dipole response (md). Electric field inside the particle at this resonance wavelength has a ring shape while magnetic field oscillates in the particle center (see Fig.1b and Supplementary Figure 2). Magnetic dipole resonance is the only peak observed for the smallest nanoparticles (Fig.3a). At increased nanoparticle size (Fig.3b&c) electric dipole (ed) resonance also appears at the blue part of the spectra, while magnetic dipole shifts to the red. For relatively small nanoparticles, the observed colour is mostly defined by the strongest resonance peak and changes from blue to green, yellow, and red when magnetic resonance wavelength shifts from 480 nm to 700 nm (Fig.3a-d). So, we can conclude that the beautiful colours observed in the dark field microscope (Fig.2a) correspond to magnetic dipole scattering of the silicon nanoparticles, "magnetic light". Further increase of the nanoparticle size leads to the shift of magnetic and electric dipole resonances further to the red and infra-red frequencies, while higher multipole modes such as magnetic and electric quadrupoles appear in the blue part of the spectra (Fig.3d-f).

Though the similarities between experimental and theoretical spectra in Fig.3 are clearly visible, there are some important differences in the spectral locations and strengths of particular resonances, which should be discussed. These differences obviously appear due to the presence of silicon substrate, which is not taken into account in the simple Mie theory solution shown in Fig.3 (iv). To study how the substrate can affect the observed spectra an additional theoretical analysis has been done. We limit this analysis to the dipole approximation, according to which each particle can be considered as point-like electric and magnetic dipoles located at the particle's center. This approximation works well for smaller particles (Fig.3a-c) when higher multipole modes are not yet excited. For further consideration we choose particle #2 (Fig.3b) as a model system, which has both electric and magnetic dipoles in the scattering spectrum while higher-order multipole modes are absent (Fig.4).

The detailed analysis of the radiation pattern of horizontal electric and magnetic dipoles above a high-permittivity substrate reveals that magnetic dipole radiates mostly to free space, while electric dipole radiates mostly into the substrate [51]. Moreover, the red shift of the measured electric dipole resonance (with respect to the Mie theory solution in free space) can be accounted for by the "surface dressing effect", described in Refs. [52, 53], according to which the electric polarizability of the sphere should be modified in the presence of the substrate. Similar approach has been suggested in Ref. [54] where the problem of the scattering by a particle on a substrate was mapped onto the problem of the scattering by a particle in a homogeneous surrounding medium with an effective permittivity. The exact value of this effective permittivity varies for different multipole modes [54]. To justify these arguments we have theoretically calculated the backward radiated power of induced electric and magnetic dipoles over a substrate based on the dyadic Green's function approach [55] with the "dressing effect" being taken into account. The results are shown in Fig. 4 and clearly demonstrate that the electrical resonance is red-shifted and significantly suppressed while magnetic dipole resonance is almost unaffected which is in very good agreement with our experimental measurements. As a result of this additional analysis one can see that the Mie theory solution in free space can provide correct assignments of different multipole resonances observed experimentally. However, for more accurate determination of the resonance maximum positions and their relative strength the influence of substrate should be taken into account.

In conclusion, strong magnetic dipole response of spherical silicon nanoparticles in the visible spectral range has been demonstrated experimentally for the first time. This resonance arises due to excitation of a particular mode inside the particles with circular displacement current of the electric field and magnetic field oscillating up and down in the middle of the particle similar to "magnetic dipole". The spectral position of this strong resonance can be tuned throughout the whole visible spectral range from violet to red by changing the nanoparticle size in the range from 100 to 200 nm.

In this work, Si nanoparticles of various sizes have been fabricated by laser ablation and studied one by one by optical microscopy, single nanoparticle dark-field scattering spectroscopy, and scanning electron microscopy. However, the use of more sophisticated methods such as laser induced transfer [56-58] can allow fabrication of nanoparticles with controlled size and precise spherical shape, and assemble them in different structures. This will open new opportunities for design of novel low-loss visible-range metamaterials and nanophotonic devices based on dielectric nanoparticles.

**Methods**

Si nanoparticles of various sizes have been fabricated by femtosecond laser ablation of a silicon wafer. In these experiments, we used a commercial 1 kHz femtosecond laser system (Tsunami+Spitfire, Spectra Physics) delivering 1 mJ, 100 fs laser pulses at a central wavelength of 800 nm. The laser beam with diameter 4mm was focused onto the sample surface by a 20x microscope objective (Mitutoyo, MPlan NIR 20). To ablate a moderate amount of nanoparticles, high enough to have all the required sizes but small enough to avoid an overlap contribution from different particles, the laser beam was scanned through the sample surface at average power of 0.25 mW and scanning speed of 1 mm/s. These parameters are slightly above the ablation threshold of silicon at these irradiation conditions. Scanned lines were separated by a pitch of 100 μm to avoid overlapping of ablated particles from different scans. The ablated particles were observed as bright color spots in a dark-filed microscope (Nikon, Ti-U) at 100x magnification. Particles with different colors were selected and analyzed by single nanoparticle dark-field scattering spectroscopy combined with the microscope and scanning electron microscopy (SEM,

JOEL 7401F). The experimental scattering spectra were compared to analytical solution of Mie theory [50], calculated using actual particle sizes measured by the SEM.

Optical scattering and spectroscopic properties of single nanoparticles were explored with an implemented high-sensitivity multifunctional microscopic and spectroscopic platform. In this platform, the white light from the halogen lamp of an inverted microscope (Nikon Ti-U) was focused by a dark-field objective (Nikon NA0.8, 100×) before being applied to excite the nanoparticles on the sample. The reflective scattering light was collected with the same objective lens and directed into the spectrometer (Andor SR-303i) with an 400 × 1600 pixel EMCCD (Andor Newton) detector attached to the microscope. A color camera was also attached to another port of the microscope to collect optical images of the scattering pattern in the dark-field illumination mode. The scattering spectra were measured over the wavelength range of 400-900 nm.

To theoretically analyze experimental data we employed Mie theory for light scattering by spherical particles in free space [50]. According to this theory we were able to identify all observed resonances based on scattering coefficients. The role of the substrate in the light scattering by nanoparticles has been investigated in the quasi-static limit based on Fourier integral representation of the dyadic Green's function method and image approach [55]. This method can be applied to both electric and magnetic dipoles in order to find their total scattered power in the far field. To correctly predict the far field profiles based on this method, the lateral directions, i.e. the regions very close to the interfaces, should be excluded [55], which exactly corresponds to the experimental setup in our case. Moreover, according to Ref. [52] the electric polarizability of the sphere $\alpha_E$ is modified in the presence of the substrate: $\alpha_E = \alpha_E^0 / (1 - \alpha_E^0 r^p / 4\pi(2h)^3)$, where $\alpha_E^0$ is the electric dipole polarizability of the sphere in free space, $r^p$ is the Fresnel scattering coefficient of p-polarized light, and $h$ is the location of the electric dipole above the substrate. This "dressing effect" results in the redshift of the electric dipole resonance and gives an excellent agreement with the experimental data.

ACKNOWLEDGMENTS

Dr. Andrey Evlyukhin from Laser Zentrum Hannover is acknowledged for stimulating discussions. The authors also thank Dr. Guillaume Vienne and Dr. Reuben Bakker from DSI for their assistance with organization of the experiments, and Ms. Janaki DO Shanmugam for help with SEM analysis. This work was supported by the Agency for Science, Technology and Research (A*STAR) of Singapore: SERC Metamaterials Program on Superlens, grant no. 092 154 0099; SERC grant no. TSRP-1021520018; the grant no. JCOAG03-FG04-2009 from the Joint Council of A*STAR, and by the Australian Research Council through the Future Fellowship project FT110100037.

AUTHOR CONTRIBUTIONS

AIK fabricated silicon nanoparticles, contributed to spectral and SEM analysis, and wrote the paper; AEM performed the theoretical analysis and contributed to the manuscript preparation; YHF and JBZ implemented the microscopic and spectroscopic platform and supervised the spectral measurements; YHF performed FDTD simulations of electromagnetic field distribution inside the nanoparticles, and carried out single nanoparticle dark-field spectroscopic measurements; BL supervised the whole work and contributed to the manuscript preparation. All authors read and corrected the manuscript before the submission.

**Figure legends**

**FIGURE 1.** Schematic representation of electric and magnetic field distribution inside a metallic split-ring resonator **(a)** and a high-refractive index dielectric nanoparticle **(b)** at magnetic resonance wavelength.

**FIGURE 2.** Dark-field microscope **(a)** and top-view scanning electron microscope (SEM) **(b)** images of the same area on a silicon wafer ablated by a femtosecond laser. Microscope image is inverted in horizontal direction relative to that of the SEM. Selected nanoparticles are marked by corresponding numbers 1 to 6 in both figures.

**FIGURE 3.** Close-view dark-field microscope (i) and SEM (ii) images of the single nanoparticles selected in Fig.2. Figures 3 (a) to (f) correspond to nanoparticles 1 to 6 from Fig.2 respectively. (iii) Experimental dark-field scattering spectra of the nanoparticles. (iv) Theoretical scattering and extinction spectra calculated by Mie theory for spherical silicon nanoparticles of different sizes in free space. Corresponding nanoparticle sizes are defined from the SEM images (ii) and noted in each figure.

**FIGURE 4.** Influence of silicon substrate on nanoparticle scattering spectra: (black curve) Experimental dark-field scattering spectrum of particle #2 taken from Fig.3b; (blue curve) Calculated scattering efficiency of a silicon nanoparticle with size of 138 nm in free space using Mie theory; and (red curve) Dyadic Green's function approach for electric and magnetic dipole radiation over the substrate with the use of "dressing effect".

**SUPPLEMENTARY FIGURE 1.**

Hierarchy of the electromagnetic resonances of spherical high-refractive index dielectric particles calculated in the frame of Mie theory. This figure demonstrates that above certain value of the refractive index the position of all multipole resonances corresponds to a fixed ratio of the wavelength inside the particle to its geometrical radius. In the figure: $R$ is the nanoparticle radius, $n$ is the refractive index of the nanoparticle dielectric material, $\lambda$ is the wavelength of light.

**SUPPLEMENTARY FIGURE 2.**

FDTD simulations of electric field distribution inside and outside a silicon particle with diameter 140 nm (Fig.3b) in free space irradiated by a plane wave from top. The plots show field distributions in the particle centre at (a) electric (476 nm) and (b) magnetic (580 nm) resonance wavelengths. The white arrows show direction of displacement currents inside the particle. It can be seen that at 580 nm electric field/displacement currents have a ring shape which leads to the magnetic dipole radiation.

**Fig.1**

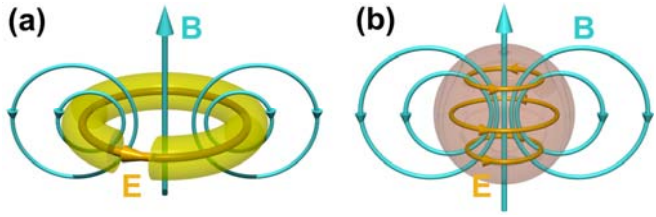

**Fig.2**

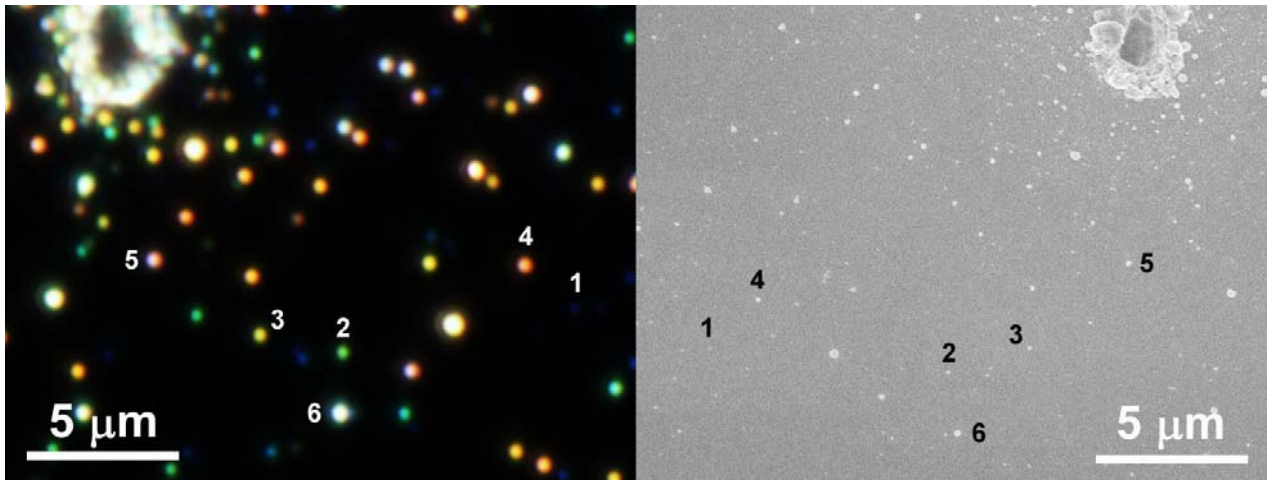

**Fig.3**

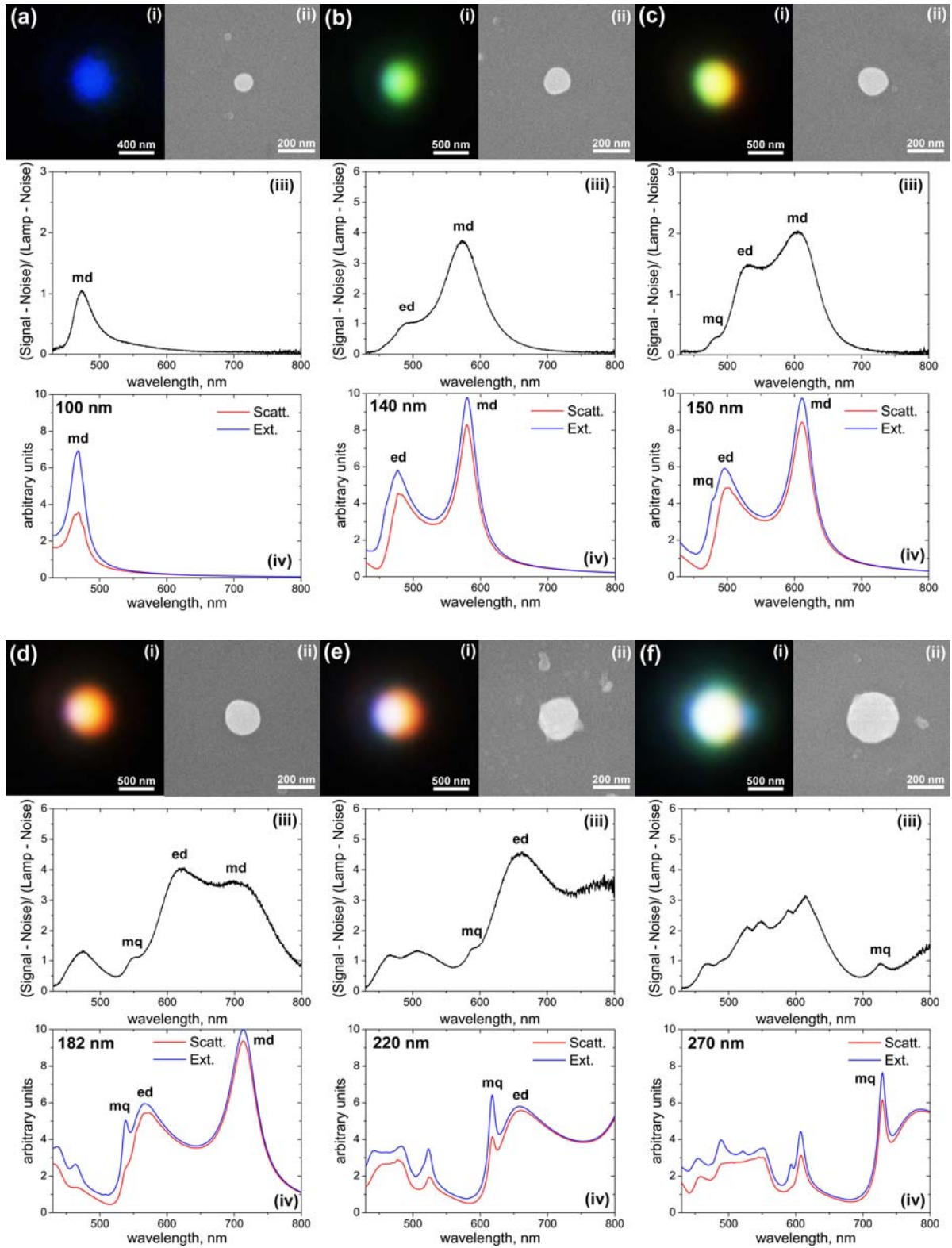

**Fig.4**

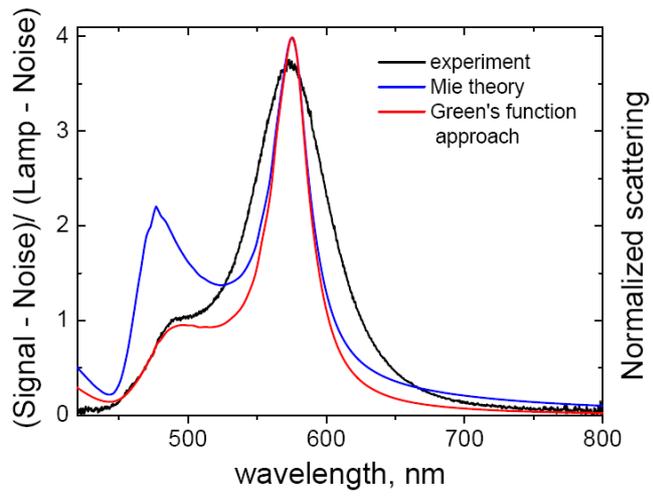

**Supplementary Figure 1**

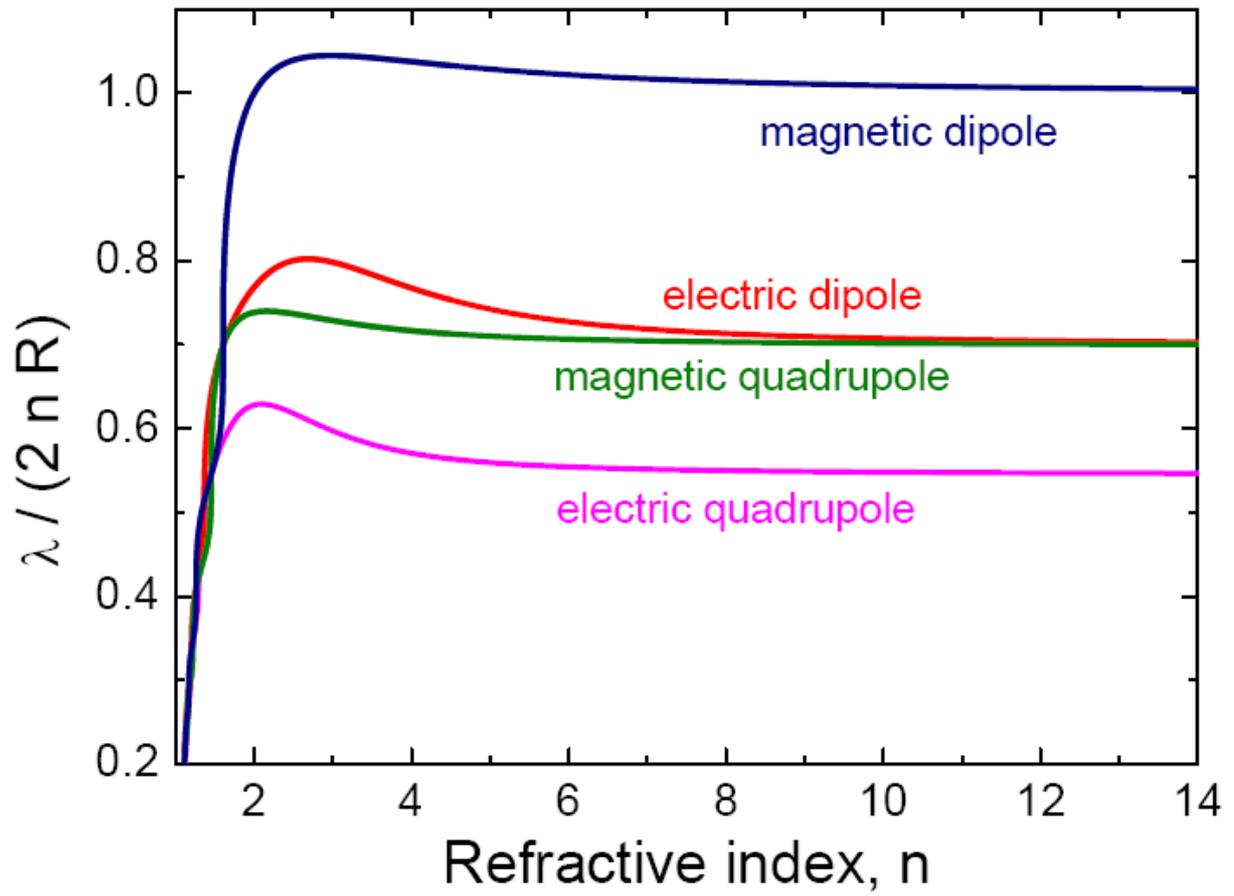

**Supplementary Figure 2**

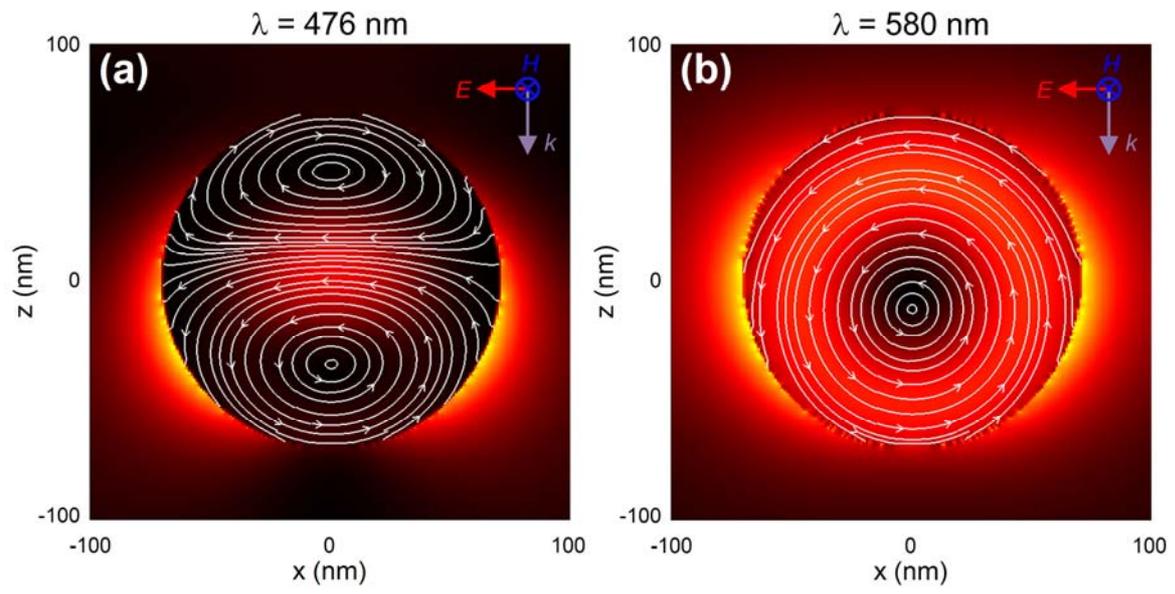